\documentclass[conference,usenatbib]{basi}
\usepackage[T1]{fontenc}
\usepackage[british]{babel}
\usepackage[varg]{txfonts}
\usepackage{rotating}
\usepackage{dcolumn}
\begin{document}
\title[Disk model in the Golden Jubilee of X-ray sources]{Status of the accretion flow solution
in the Golden Jubilee year of the discovery of extra-solar X-ray Sources}
\author[S.~K.~Chakrabarti]%
       {S.~K.~Chakrabarti$^1$\thanks{email: \texttt{chakraba@bose.res.in}}\\
       $^1$S. N. Bose National Centre for Basic Sciences, JD Block, Salt Lake, Kolkata, 700098, India \\
       $^2$ Indian Centre for Space Physics, 43 Chalantika, Garia Station Rd., Kolkata, 700084, India}

\pubyear{2013}
\volume{**}
\pagerange{**--**}

\date{Received --- ; accepted ---}

\maketitle
\label{firstpage}

\begin{abstract}
Fifty years have just passed since the first discovery of the extra-solar
X-ray sources by Giacconi  and his team \citep{Gi67} which we know today to be some stellar mass black holes. 
By 1973, not only a catalog of these enigmatic objects were made, and their spectra were obtained.
Today, forty years have passed since the revolutionary idea of the thin, 
axisymmetric, Keplerian, disk model by Shakura and Sunyaev was published. 
Yet, the complete predictability of their radiative properties remains as illusive 
as ever. The only available and self-consistent solution to date is the generalized 
viscous transonic flow solutions where both heating and cooling effects are included. I demonstrate that the   
latest `Avatar' of the accretion/outflow picture, the Generalized Two Component Advective Flow (GTCAF),
is capable of explaining almost all the black hole observational results, when the results of the
time dependent simulation of viscous and radiative processes are also 
taken into consideration. I also discuss the problems with predictability and argue that 
understanding companion's behaviour in terms of its habit of mass loss, 
ellipticity of its orbit, magnetic properties, etc. is extremely important 
for the prediction of emission properties of the accretion flow. 
\\[6pt]
\hbox to 30pt{\hfil}\verb|http://www.ncra.tifr.res.in/~basi/|
\end{abstract}

This paper is to be cited as: S.K. Chakrabarti, Bulletin of Astronomical Society of India,
Proceedings of RETCO conference, S. Das, A. Nandi and I. Chattopadhay (Eds.) 2013.


\section{Introduction}\label{s:intro}

Fifty years ago, the subjects of the physics of stellar mass 
black holes and the Super-Massive black holes started
(e.g., \citep{Gi67, Ma73, Sc71, BBS73} and references therein) 
By 1973, the thin accretion disk model of
\citep[][Hereafter SS73]{SS73} and \citep{NT73} was proposed, not only to explain the soft-Xray 
bumps in stellar systems, but also the big-blue-bump in ultra-violet region in Quasars and Active Galaxies.
The euphoria that the subject is almost closed disappeared soon with the 
discoveries of instabilities associated with the SS73 disks \citep{EL74} and with power-law high energy X-rays 
in stellar mass objects (\citep{ST79, C96a, C08} and references 
therein). With the launching of the RXTE satellite,
the time variation of spectral and timing properties of several black 
hole candidates were revealed and it became
clear that the focus must be to understand the accreting system as a whole,
and not parametrically on a case by case basis. 
The solutions of governing equations which make minimal assumptions
survive the test of time and thus one requires to sharpen these 
theoretical solutions having maximum possible flexibility, while still remaining 
the solutions of the governing equations. It is with this spirit, the two component 
advective flow (TCAF) soilution  by \citep[][hereafter CT95]{CT95}
was presented to the community which combines the theory of viscous transonic 
flow \citep{C90a} with the radiative properties of an electron 
cloud \citep{ST80,ST85}. 

Though it was known that the flows on a black hole should be transonic in nature, 
\citep{C89,C90a} firmly established
that if one wants to completely explain the black hole astrophysics, i.e., both
spectral and temporal properties,  one must live with an 
ugly feature of the transonic accretion flow -- a major part of the parameter space 
spanned by energy, angular momentum and viscosity, would force the flow to have a centrifugal
pressure supported shock wave inside the disk. Unfortunately, this fact is difficult to digest
by minds which, albeit irrationally, believe in simpler pictures, very much like workers
who believe in a flat earth or a steady state Universe even in the presence of 
inundated evidences to the contrary. To make the
matter worse, Chakrabarti and his collaborators \citep[][hereafter MSC96]{MSC96}, \citep[][hereafter CAM04]{CAM04} and
\citep[][hereafter GGC13]{GGC13} firmly established
that when the solution is allowed to be time-dependent
the parameter space is further extended to include oscillating shocks in the disk. Explanation
of both the spectral and temporal properties by a single theory, which 
is inevitable in any subject, thus became a reality \citep{D13a, D13b}. We are thus truly
looking at the light at the end of the long tunnel that opened half a 
century ago. Everyday, evidences are piling up that the accretion flow indeed 
has two active components e.g., \citep{Sm01,Sm02,Sm07,CS13}.

This present brief review is to support these tall claims with some
examples. In the next Section, I describe the building blocks of 
a generalized transonic flow which have been rigorously established. In Section 3, we give a summary of the 
flow behaviour. In Section 4, I present a discussion of predictability of the TCAF solution.
Finally, in Section 5, we draw our conclusions.

\section{Building Blocks of Generalized Transonic Flows around a black hole} 

Here are a few major building blocks which are responsible for 
the observed spectral and temporal properties of the flow. We briefly discuss them.

\subsection{The Existence of a Critical Viscosity}

The most important of all is the discovery that there exists a critical viscosity above which the 
solution will not allow a steady shock wave in the flow and below that, a 
standing shock wave would be present. Figure 1a shows the 
Mach number (Y-axis) vs. radial distance (X-axis) for a viscous isothermal flow
for illustration \citep{C90b}. The inner sonic point ($X_{in}$), the sound speed ($K$) and the 
SS73 viscosity parameter $\alpha$ are marked inside the Figure. Here, the solution passes through both the inner and the
outer sonic points. This viscosity parameter $\alpha_{in}$ is the critical parameter in this case.
When $\alpha <\alpha_{in}$, the flow has a shock wave solution. For 
$\alpha > \alpha_{in}$ the flow simply becomes a Keplerian disk entering into a black hole
through the inner sonic point. Details of the critical parameters for other models 
may be seen in \citep{C96b,CD04,DC04}. The numerical simulations 
to demonstrate this is in \citep{LMC98,GC12}. 

\begin{figure}
\centerline{\includegraphics[width=5cm]{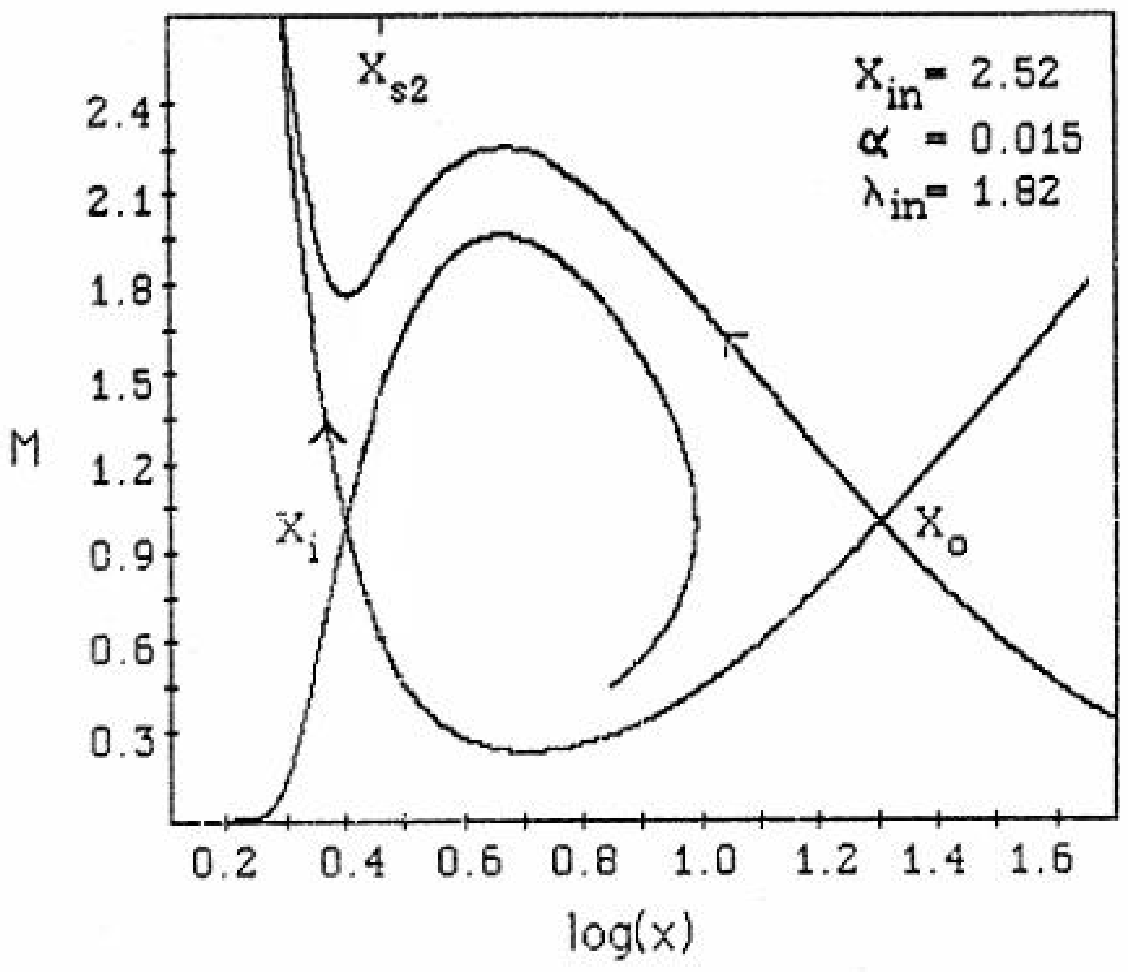} \includegraphics[width=5cm]{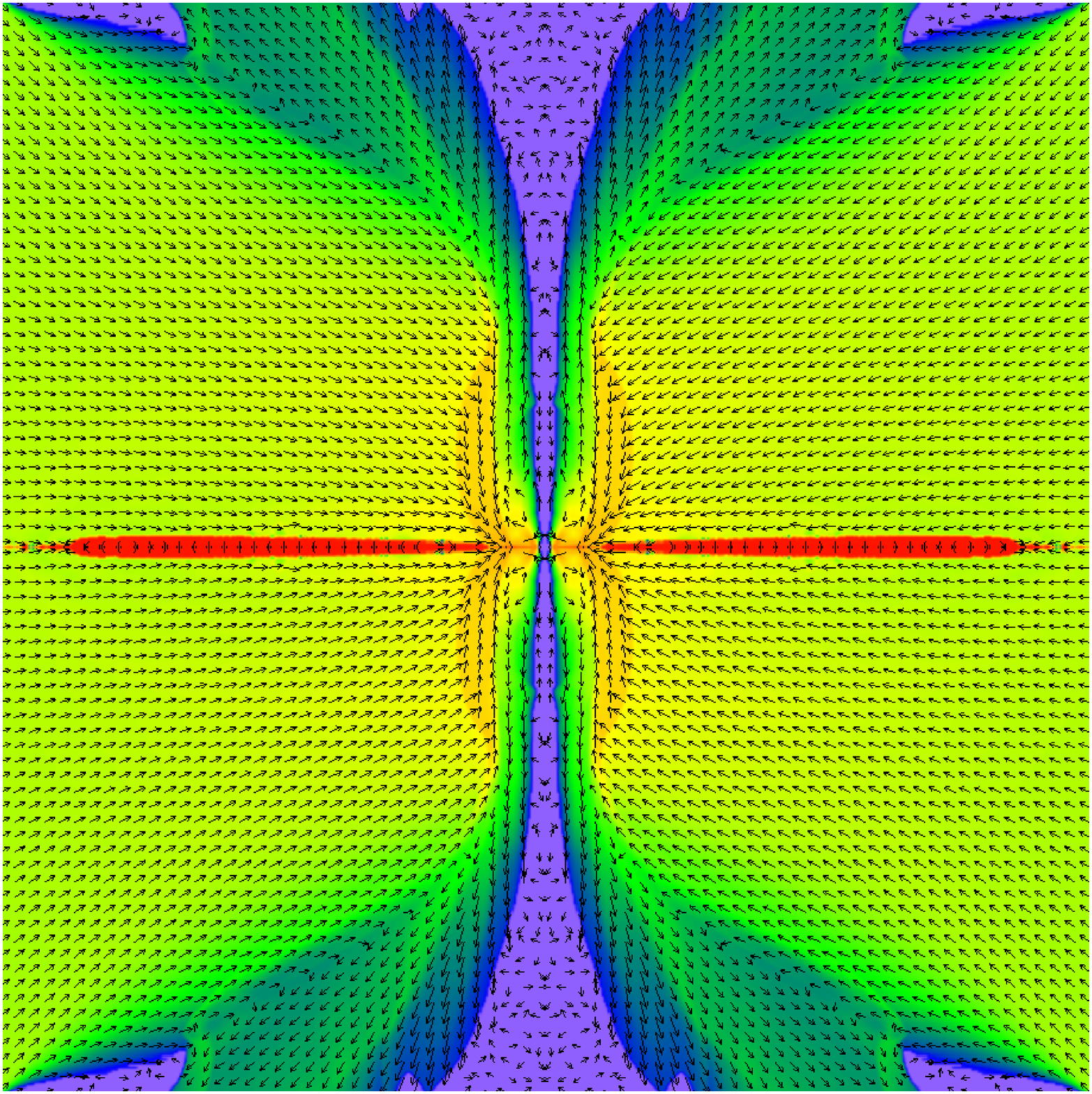}}
\caption{(a) (left) An example of the topology with the critical viscosity parameter $\alpha_{in}$ 
which separates the parameter space. For $\alpha<\alpha_{in}$ the flow can have shocks
and for $\alpha>\alpha_{in}$ the flow will be a Keplerian disk passing through the inner sonic point
\citep{C90a}. In (b), (right) we show the formation of TCAF with $\alpha>\alpha_{in}$ Keplerian disk on the equatorial 
plane, and $\alpha<\alpha_{in}$ sub-Keplerian halo away from the equatorial plane (Giri \& Chakrabarti, 2013).}
\end{figure}

A ramification of this critical viscosity is that when the black hole
accretes a low-angular momentum flow and there is a sudden surge of viscosity 
the flow quickly transports angular momentum, and enters into the black hole as a Keplerian disk. However,
above a certain height from the equatorial plane, the flow may still
have a lower viscosity, and thus will produce a shock wave. In other words,
the flow would be divided into two components (CT95, \citep{C96b,C97}). This is the 
well studied Two Component Advective Flow (TCAF) solution. Recently, \citep[][Hereafter GC13]{GC13} showed that this 
configuration is indeed stable. Thus the spectral studies of CT95 are realistic.
Figure 1b shows the configuration formed by GC13
through numerical simulations. The equatorial 
plane has a Keplerian disk and the sub-Keplerian halo surrounds it. 
CENtrifugal pressure supported BOundary Layer or CENBOL is also formed. 

\subsection {The Existence of a Resonance Oscillation of the Shocks}

MSC96 and CAM04, using power law cooling, and more recently, GGC13, using Compton cooling
showed that when the infall timescale from the post-shock flow `roughly' matches with the cooling time scale, the shock oscillates
with a frequency which is inverse of the infall time scale from the shock to the inner sonic point.
Oscillation of the shock wave may occur when the Rankine-Hugoniot condition is not fulfilled, and yet
ithere are two physical sonic points in the flow \citep{RCM97}. While
these oscillations are robust, more studies are to be carried out for other cooling effects
(such as synchrotron radiation), non-axisymmetric flows etc. \citep{C06}.

\subsection{The outflows originating from post-shock region}

A quantitative measurement of the outflow rate was made by \citep{MLC94}.  
Later, in \citep{C08,DC99,DCNC01,SC11}
the ratio of outflow to inflow rate was obtained by purely
theoretical means. It was observed that a few percent of the inflow leaves the system
as an outflow and hotter the post-shock region is, or higher the angular momentum is,
higher is the ratio. This outflow was found to be partly or totally 
iquenched as the accretion rate of the Keplerian rate goes up \citep{GGC13}

\section{The Generalized TCAF solution}

Figure 2 shows schematically all the components of a generalized TCAF solution (GTCAF)
which traces most of the matters' journey from the companion to the black hole and the rest 
towards forming the jets. The high viscous Keplerian disk is immersed inside a low angular momentum halo component.
For all practical purposes, CENBOL is the boundary layer of a black hole, except that 
instead of being supported by a hard surface, it is supported by a centrifugal barrier. The steady jets and 
outflows in this picture are produced primarily from CENBOL. The CENBOL and the base of 
the Jet (pre-Jet) is also the so-called Compton cloud which intercepts soft-photons 
from the Keplerian disk and re-radiates them after Comptonization. The jets are 
accelerated by combinations of the hydrodynamic, magnetic and radiative forces 
\citep{CDC04}. The collimation is possibly done by the 
toroidal field lines which are ejected from the disks due to buoyancy effects \citep{C94}. 
The disks may not have a constant accretion rates at all radii, especially in the Keplerian disk, 
since the matter supply from the companion may vary in the orbital and viscous time scales. 
\begin{figure}
\centerline{\includegraphics[width=9.0cm]{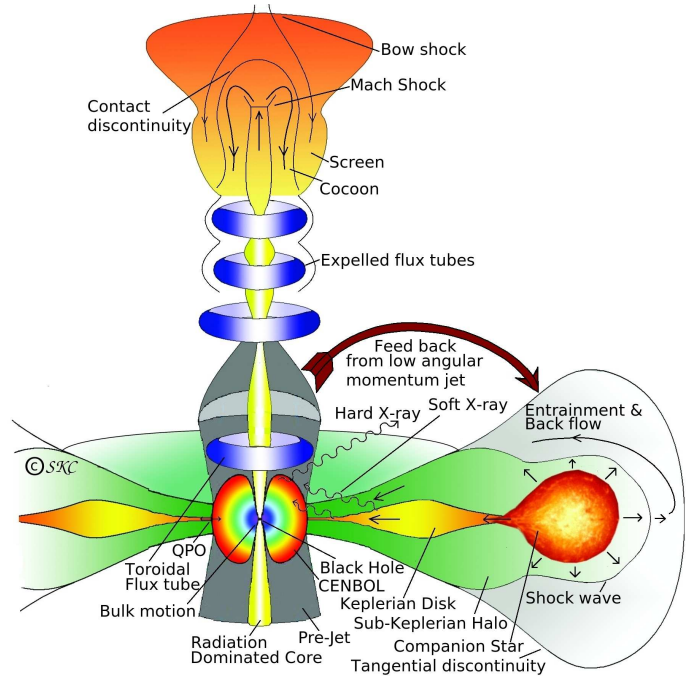}}
\caption{Generalized TCAF with CENBOL, Keplerian and sub-Keplerian flows and the pre-Jet configuration
gives the whole scenario which includes the companion star and double 
jets. The schematic diagram is roughly in logarithmic scale in both directions 
and different relevant components are zoomed in to show details. } 
\end{figure}

In the case of outburst sources, fitting of the data with a TCAF solution suggests that the main cause could be the 
sudden rise in the viscosity at the outer boundary (perhaps due to magnetic activities of the companion star, matter accumulation 
at the outer edges). As long as the viscosity is high enough, the rising phase persists. 
During the intermediate phases \citep{N12,D13c}, the viscosity 
remains moderately high. Declining phase starts only when viscosity starts to decrease.

In the TCAF, the low angular momentum matter may be supplied in many ways: First of 
all, the stars will have winds which may produce shocks all around to 
satisfy the boundary condition that the ambient medium is stationary. These shocked flow could be be directed to 
the central black hole. Since they are coming from spherically symmetric (in the frame of the 
companion) flow, they would have very low angular momentum. Another possibility is to have a part of subsonic
pre-jet and outflow (which are formed out of very low angular momentum of the inner CENBOL region) 
returns back to the equatorial plane at a larger radii and falls back to the black hole. Similarly, magnetic field of the 
companion, if entangled with the outer disk, could not only remove angular momentum from a Keplerian disk there,
it would bring sub-Keplerian matter from the stellar pole to the outer disk. It requires a 
complete numerical simulation to establish the complete scenario.

\section{Predictability of TCAF Solution}

Since TCAF is the more general solution, in the appropriate limit, it will have the same predictability as those 
of the Shakura-Sunyaev disk (SS73), or a thick accretion disk 
or a radiatively inefficient transonic flows \citep{C89}. No solution is expected to have a predictability 
more than  a fraction of the orbital time of the companion. TCAF requires outer boundary condition 
and these are not properly known. However, given such boundary values of disk and halo rates, TCAF is capable of predicting 
several observations. We present some examples due to sort of space.

\noindent A. Spectral state transition

In a TCAF, the relative importance of the soft photon source (accretion rate of the Keplerian 
disk) and the hot electrons (rate of the low angular momentum matter) decides whether the 
CENBOL could be cooled or not. If yes, it would be a spectrally soft state, otherwise it would be a
spectrally hard state. In CT95 and \citep{E96} paper, it was shown
that the energy spectral index $\alpha$ in hard state is insensitive to the optical depth of the CENBOL. 
At a disk rate of around Eddington rate, the sensitivity $\alpha$ on disk rate
is very high and the transition to soft state is very rapid. Meanwhile, as CT95 showed, the 
spectral index due to the Bulk Motion Comptonization is dominating. There is a regime
when broken power law spectra could be seen, though, at a very high disk rate (${\dot m}_d >>1$)
the spectrum will have a power-law slope corresponding to the bulk motion. Spectra of 
Cyg X-1 is fitted very well \citep{CM06} when synchrotron  
photons along with shock accelerated electrons also participate in the process. Similarly,
the spectra of supermassive black hole M87 also fitted well using TCAF solution \citep{MC08}.
So far, TCAF has been implemented in XSPEC without taking the bulk motion Comptonization (BMC) into account. 
Hence better fits are obtained for harder states \citep{D13a,D13b}. The BMC is being 
incorporated. Fitting of the spectral data is possible with the TCAF solution. So far, TCAF has been implemented in XSPEC 
without taking the bulk motion Comptonization into account. Hence, better fits are obtained for harder states.
Five parameters, namely, the accretion rates of the disk, halo, shock location, compression ratio and the 
mass of the black hole are extracted out of the XSPEC fit. TCAF is the only solution in which spectral 
fits yield temporal behaviour, since the shocks which fit the spectra also produce QPOs. For details,
see \citep{D13a,D13b}

\noindent B. Energy dependence of QPO behaviour

Since the TCAF solution shows that QPOs are formed due to the oscillation of the Compton cloud
(CENBOL), only Comptonized photons are supposed to take part in QPOs. Thus, TCAF predicts 
that the power in QPO oscillation should be very low for the photons coming out of a Keplerian
disk, but the power would be higher for the photons which are Comptonized. 
However, too many scatterings will increase the energy but will reduce coherency of 
oscillations \citep{CM00}
\noindent C. QPO frequency and resonance condition

Since the CENBOL cools down and moves in when the Keplerian rate is increased, the QPO frequency 
should go up accordingly (MSC96, \citep{CM00,C08,D13c}). Typically, 
$$
\nu_{QPO} \sim t_{infall}^{-1} \sim \frac{\sqrt{G M_{BH}}}{R f_t (X_s)^{3/2}} {\rm  Hz} ,
\eqno{(1)}
$$
where, $X_s$ is the shock location in Schwarzschild radius, $R$ is the compression
ratio (ratio of post shock density and pre-shock density), $G$ is the gravitational 
constant and $M_{BH}$ is the mass of the black hole. This can be rewritten as,
$$
\nu_{QPO}=18.5 m_{10}^{-1} f_{t,3}^{-1} r_4^{-1} X_{s,10}^{-3/2} {\rm Hz} .
\eqno{(2)}
$$
Here, $m_{10}$ is the mass of the black hole in units of $10 M_\odot$, $r_4$ is the compression
ratio in units of $4$, $X_{s,10}$ is the shock location in units of $10r_g$. Here, we have introduced a
factor $f_t$ which reduces the infall velocity due to the
centrifugal pressure induced turbulence inside the CENBOL. A reasonable guess would be $f_t \sim \lambda^2$,
where $\lambda$ is the dimensionless angular momentum. $f_{t,3}$ is in units of $3$. 
The QPO frequency goes up as the shock location goes down.

After making simplified assumptions, and computing cooling time scale assuming a constant enhancement
factor ${\cal E}=20$, injected soft photon energy $e=0.5$keV and the shock to be in vertical 
equilibrium, one can show that the ratio of the cooling and the infall time scale is, 
$$
\frac{t_c}{t_{infall}} = 0.4 {\dot m}_{18}^{1/4}  X_{s,10}^{-9/4} {\cal E}_{20}^{-1} e_{0.5}^{-1}.
\eqno{(3)}
$$
Here, we chose the mass the black hole to be $10M_\odot$, $\gamma=4/3$ to be the adiabatic index
of the flow, $R=4$ is the compression ratio of the gas at the shock. ${\cal E}_{20}$ is the 
average factor by which the injected photon energy is enhanced on an average as is in units of $20$.
It can vary from $E_{20} \sim 1$ (very soft state), to $\sim 40$ (very hard state). Either way, when the 
ratio is off balanced from $\sim 1$ and the low frequency QPOs would not be seen. 
This little exercise shows the power of TCAF solution. It directly shows that timing 
properties are spectral properties are connected through the shock location. It also predicts 
how QPO frequency would become higher on a daily basis as the shock location 
moves towards the black hole in a outburst source (Eq. 1). For AGNs, the ratio will (Eq. 3) will not be 
close to 1 when Compton cooling is used. In that case bremsstrahlung cooling could cause QPOs (MSC96).
Figure 3, taken from GGC13 shows how the QPO frequency increases with  
the accretion rate of the Keplerian disk. This behaviour is observed in the rising phase of the
outburst sources. Indeed, the fit of QPO frequency with time is acceptable enough so that from 
the first three to four days of data, one could predict the QPO frequency variations
in the next several days -- which is impossible by any other model.      

\begin{figure}
\centerline{\includegraphics[width=6cm]{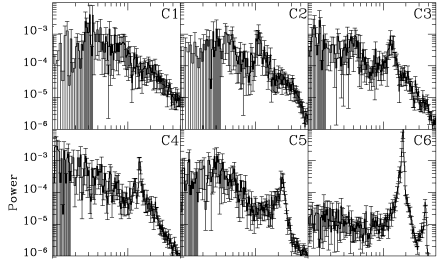}}
\caption{Demonstration of increasing QPO frequencies when the disk rate is increased. This result 
comes totally from a time dependent hydrodynamic simulation coupled with the Monte Carlo simulation of Comptonization.
See, GGC13 for details. }
\end{figure}

\section{Concluding Remarks} 

In this article, we presented briefly the  current status of solution onto black holes. It is to be noted that 
this solution was presented {\it before} the RXTE was even launched. It turns out to be 
very promising, as it can satisfactorily explain the state transitions
as far as the spectral properties go, and timing properties such as low frequency QPOs 
and their time variation during the outbursts. To our knowledge, no other solution has 
as much predictability. However, as we pointed out, the biggest problem is the poor knowledge of 
the boundary condition for solving the differential equations which govern the
flow. As shown in Fig. 2, ideally, one needs to treat ``matter supply - accretion flow - jet"  
as a single physical system. However, the companion's property could be erratic. The mass transfer rate 
could vary with phase, stellar magnetic cycle, X-ray irradiation, inclination of 
the companion's spin axis etc. If the orbit is elliptic this should be imprinted on long term count rate variation.
Entangled magnetic field may also remove angular momentum from the accretion disk, supplying the halo with low angular momentum 
matter. Nevertheless, we can assume that in a short time span (as compared
to the orbital time of the binary), the conditions at the outer boundary may remain the same 
(rate and its derivative). Thus we may extrapolate the results successfully using TCAF in near future and near past of an observed event. 
Because of this, the variability class transitions of GRS 1915+105 is also very much unpredictable.
Fortunately, one parameter which is defined to be the Comptonization efficiency (CE)
(ratio of the power-law photons number and the black body photon number) has been found which mixes the 
properties of the two flow components. When Compton efficiencies are plotted against
variability classes so that observed transitions take place among nearest neighbours, CE becomes a 
monotonically increasing straight line, lowest CE being for the softest class \citep{PNC13}.
This CE is the dynamic hardness ratio valid for all black hole candidates.

\section*{Acknowledgements}

I thank all my students and collaborators who have helped completing various aspects of the TCAF 
solution.

\end{document}